\documentclass[aps,pra,twocolumn,showpacs,superscriptaddress,groupedaddress]{revtex4}  % for review and submission
\usepackage{graphicx}  % needed for figures
\usepackage{dcolumn}   % needed for some tables
\usepackage{bm}        % for math
\usepackage{amssymb}   % for math

\begin{document}

% The following information is for internal review, please remove them for submission

% the following line is for submission, including submission to the arXiv!!
%\hspace{5.2in} \mbox{Fermilab-Pub-04/xxx-E}

\title{Higher entanglement and fidelity with a uniformly accelerated partner via non-orthogonal states}
\author{G. A. White}
\affiliation{Department of Physics and Astronomy, University of Kentucky, Lexington, Kentucky 40506-0055, USA}
\affiliation{School of Physics, Monash University, Melbourne, Victoria 3800 Australia +
ARC Center of Excellence for Particle Physics at the Tera-scale, Monash University, Melbourne, Victoria 3800 Australia. Email: graham.white@monash.edu}

\date{\today}
\begin{abstract}

We show that nonorthogonal states achieve a higher level of entanglement when one party undergoes uniform acceleration. We also show that in this regime non-orthogonal states achieve a higher fidelity for teleportation. A quantum field as observed by an observer confined to a Rindler wedge acts as an open system. When one traces over the field modes in the causally  disconnected region of space time, one finds that,  for a qubit derived from the modes of a scalar field, the effective state space of a qubit is deformed. In order to show this we numerically calculate the Bures angle between states as seen by an accelerating observer. We also derive the Bures metric of an effective qubit in the limit of low accelerations. In order to show that the deformation is not just a coordinate transform, we calculate the scalar curvature of the effective Bures metric. We see that the effective Bures metric loses its isotropic nature.  Despite taking the limit of low accelerations, our results are also relevant to large gravitational fields.

\end{abstract}

\pacs{03.65.Bz, 03.67.Hk, 03.30.p, 03.650d, 04.62.+v}
\maketitle

%\section{\label{sec:level1}First-level heading}
% sections are not used for PRL papers
\section{Introduction}
A plethora of practical applications have arisen due to the study of quantum information theory. This study has also produced equally interesting insights into physics questions of a fundamental nature. Of particular relevance to this paper is the recent progress made in the study of quantum information theory in relativistic settings \cite{first}. When an observer undergoes uniform acceleration, the appropriate coordinate system to use is known as Rindler coordinates. This coordinate system contains two causally disconnected domains. The Minkowski vacuum of the field observed by our accelerated observer will be populated with particles. This phenomenon, known as the Unruh effect, demonstrates the relative nature of the vacuum \cite{unruh}. It is perhaps even more startling that the Unruh effect causes decoherence. Hence when an entangled state is observed by an accelerating observer, it appears to have less entanglement. That is, due to the Unruh effect, the amount of entanglement a state has is also relative to the observer \cite{entanglement}. Here, we demonstrate another surprising result: that the structure of the effective state space of a qubit is also relative to who is observing it.

Rindler coordinates are an appropriate coordinate system for describing an observer subject to uniform acceleration. In Minkowski coordinates, the world lines of an accelerated observer correspond to hyperbolas that asymptote to null trajectories on either side of the origin. The tangents to these asymptotes intersect at the origin and form two causally disconnected regions to the left (region I) and right (region II) of the origin \cite{unruh}. Alsing and Milburn considered this situation with two observers, Alice and Rob, each holding an optical cavity capable of supporting two orthogonal modes. Alice is a Minkowski observer and Rob undergoes uniform acceleration. At a certain event, Rob's frame is instantaneously at rest, the two cavities overlap and an entangled state is created as Rob accelerates away. Alice then attempts to teleport a state to Rob, but this process is disrupted by the Unruh effect. It is then shown that the fidelity of Rob's final state with Alice's initial state degrades due to acceleration \cite{Milburn}. However, as Schutzhold and Unruh pointed out, the presence of cavity walls block the radiation resulting from the Unruh effect and there are various other conceptual difficulties with this approach \cite{comment}. Therefore we instead consider a situation where Alice and Rob's modes are the modes of a free scalar field \cite{freefield}. The fields can be massive \cite{massive} or massless \cite{entanglement} as this will not affect the mathematical form of our results.  We are therefore making an approximation where only the main contributing mode of a wave packet is considered \cite{singlemode1,singlemode2}.  Recent work has shown that this single-mode approximation is indeed valid in certain physical situations \cite{singlemode3}. 

In Sec. \ref{setup} we explain our model of quantum information with non-orthogonal states subject to the Unruh effect. In Sec. \ref{two} we calculate the entanglement, perform a teleportation protocol and discuss distinguishability within our model. In each case we  show that there exist non-orthogonal states that outperform orthogonal states for quantum information given a particular non zero value of acceleration. We will then calculate the Bures metric of the effective qubit derived from the modes of a scalar field in the limit of low accelerations and demonstrate its deformation in Sec. \ref{three} before concluding in Sec. \ref{four}.

\section{Quantum information with non-orthogonal states and the Unruh effect}\label{setup}

In this section we will develop a simple model of quantum information with non-orthogonal states subject to the Unruh effect. The non-orthogonal basis we consider is $\{|+\rangle ,|\varphi \rangle\}$. These states can be expressed in the number occupation basis as
\begin{equation} |+\rangle = \frac{1}{\sqrt{2}}(|0\rangle + |1\rangle ) \label{plusstate} \end{equation}
and 
\begin{equation} |\varphi \rangle  = \sqrt{\frac{1-\xi}{2}} |0\rangle - \sqrt{\frac{1+\xi}{2}}|1\rangle  \label{firstequation} \ \end{equation}
respectively. Here $\xi$ is a parameter that determines the degree to which $|+\rangle$ is non-orthogonal to $|\varphi \rangle$. Note that for $\xi = 0$, $|\varphi \rangle $ is orthogonal to $|+\rangle$.  The right hand sides of Eqs. (1)--(2) are written in the number occupation basis, so it is implicitly assumed that Alice and Rob share a reference frame for phase \cite{super}.  Following the approach of other authors we ignore the specific details of the reference frame henceforth \cite{freefield}. 

As stated in the introduction, we are considering the modes of a free scalar field. The modes of a scalar field are an infinite level system from which a qubit can be prepared \cite{quantumscizzors,qubitfieldmode}. Following the approach that is common in the literature, we consider the entanglement of two qubits where Rob's qubit is subjected to the Unruh effect \cite{first,Milburn,freefield}.  As such we will encode in a non-orthogonal basis for Rob only. We therefore consider the entangled state, $| \Psi \rangle _{\rm AR}$, shared by our initially inertial observers, Alice and Rob, where
\begin{equation} |\Psi \rangle _{\rm AR} = \frac{1}{\sqrt{2}}(|+\rangle_{\rm A} |+\rangle_{\rm R} + |-\rangle_{\rm A}|\varphi \rangle _{\rm R}) \label{psipsi} \end{equation}
and \begin{equation} |\pm \rangle = \frac{1}{\sqrt{2}}(|0\rangle \pm |1 \rangle ) \ . \label{secondequation} \end{equation}

Rob then undergoes uniform acceleration, hence we must consider the quantization of a scalar field in Rindler space (which is not the same as in Minkowski space). By using Unruh's famous result we can express the Minkowski space vacuum state in terms of region I and region II modes as follows \cite{unruh}
\begin{equation} |0 \rangle _M =  \frac{1}{\cosh r} \sum _{n=0}^\infty \tanh ^n r |n \rangle _{I} |n \rangle _{ II } \label{unruh1} \ , \end{equation}
with
\begin{eqnarray} \cosh r &=& \left( 1-e^{-2\pi \Omega } \right) ^{-1/2} \ , \nonumber \\ \sinh r &=& e^{- \pi \Omega } \left( 1- e^{-2 \pi \Omega }\right) ^{-1/2}  \ , \end{eqnarray}
and regions I and II denote the two causally disconnected regions in Rindler space. Here $\Omega \equiv \omega _R /(a/c)$ where $\omega _R$ is the frequency of the Rindler particle and $a$ is its acceleration. Similarly the Minkowski space one-particle Fock state can be written in terms of the modes of region I and II as
\begin{equation} |1 \rangle _M = \frac{1}{\cosh^2 r} \sum _{n=0}^\infty \tanh ^n r \sqrt{n+1} |n+1 \rangle _{I} |n \rangle _{II} \  \label{unruh2} \end{equation}

Since the two regions are causally disconnected, the effective state of the system can be found by tracing over the field modes in region II. That is, $ \rho _I = {\rm Tr}_{II} [\rho _M]. $  Using the shorthand $\eta _{\pm \pm} \equiv 1\pm \sqrt{1 \pm \xi }$, we write our entangled state as
\begin{equation} \rho _{I {\rm AR}}= \frac{1}{8 \cosh ^2 r} \sum _{n=0}^\infty \tanh ^{2n} r \rho _n \label{effectiverho} \end{equation}
where
\begin{eqnarray} & \rho _n &= \chi \bigg[ \eta _{+-}|0,n\rangle _{\rm AR}+\eta _{--} |1,n \rangle _{\rm AR}  \nonumber \\   &+& \frac{ \eta _{-+} \sqrt{n+1}}{\cosh r}  |0,n+1\rangle _{\rm AR}+ \frac{\eta_{++}\sqrt{n+1}}{\cosh r} |1,n+1\rangle _{\rm AR}  \bigg] \nonumber  \\  \label{effectiverhon} \end{eqnarray}
and $\chi | \psi \rangle \equiv | \psi \rangle \langle \psi |$. Here and throughout this paper we omit the subscript I whenever it is convenient to do so.

\section{The superiority of certain non-orthogonal states for accelerated observers}\label{two}
In this section we demonstrate that for the model developed in Sec. \ref{setup} there exist non-orthogonal states that are superior to orthogonal states for quantum information. We remind the reader that we are comparing states that have been prepared by an inertial observer, as either orthogonal or not, and are then subject to the Unruh effect due to the observer experiencing uniform acceleration. We begin by calculating the entanglement of Eq. (\ref{effectiverho}). 
\subsection{Entanglement}
For pure states, the von Neumann entropy of Alice's (or equivalently Rob's) reduced density matrix uniquely quantifies the entanglement of a system \cite{teleportation}. However, for mixed states no such measure exists. The logarithmic negativity, however, is an entanglement monotone for which a non-zero value is a sufficient, but not necessary condition of entanglement \cite{logneg}.  Since we will only compare states with non-zero logarithmic negativity this figure of merit will suffice for our analysis. It is defined as
\begin{equation} E_N(\rho) = \log _2 \left| \left|  \rho ^ \Gamma \right| \right| \ , \label{loggyneg} \end{equation}
where $\rho ^\Gamma $ denotes the partial transpose of $\rho$ \cite{logneg}. 

\begin{figure}
\includegraphics[scale=0.43]{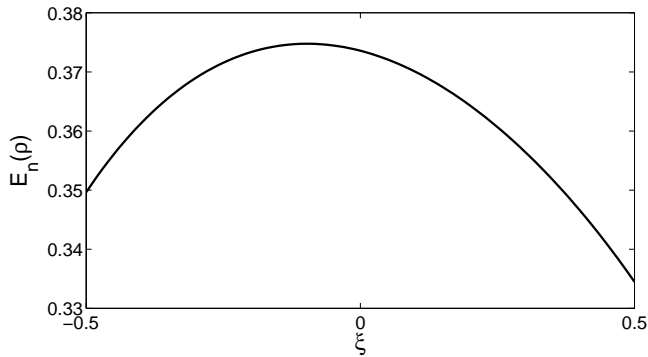}
\caption{\label{lognegfig} The logarithmic negativity of an entangled state subject to Unruh radiation as a function of our orthogonality parameter, $\xi$ ($\xi=0$ denoting an orthogonal state). The value of our acceleration parameter is $r=0.6$.}
\end{figure}
Consider the partial transpose of $\rho _n$ with respect to Alice's modes where $\rho _n$ is defined by Eq. (\ref{effectiverhon}). We can write the sections of $\rho _n $ that are invariant and non-invariant to the partial transpose as $\overline{\rho} _n$ and $\hat{\rho} _n$ respectively where $\rho=\hat{\rho}_n+\overline{\rho}_n$. The partial transpose of $\rho _n$ is then just $\overline{\rho}_n + \hat{\rho}_n^\Gamma$ where
\begin{eqnarray} \hat{\rho}_n ^\Gamma &=& \frac{\eta _{+-} \eta_{++}\sqrt{n+1}}{\cosh r}|1,n\rangle  _{\rm AR} \langle 0,n+1| \nonumber \\ &+&\frac{\eta_{--}\eta _{-+}\sqrt{n+1}}{\cosh r}|0,n\rangle  _{\rm AR} \langle 1,n+1| + {\rm H.c.} .  \nonumber \\ \end{eqnarray}
We see that the second term in the above equation is non-zero only for non-orthogonal states. In Fig. \ref{lognegfig} we show numerically that the logarithmic negativity of $\rho_{I {\rm AR}}$ as defined in Eq. (\ref{effectiverho}) can be increased using non-orthogonal states.

\subsection{Teleportation}
Here we consider Alice and Rob using their entangled state to teleport an arbitrary pure state,
\begin{equation} |\psi \rangle _{\rm Q} = \alpha |0 \rangle _{\rm Q}+ \beta | 1 \rangle   _{\rm Q} \ , \end{equation}
from Alice to Rob. To do this, Alice applies one of four positive-operator-valued measures (POVMs), $\hat{\Pi}^i_{\rm QA}$, to the joint space of her qubit and her half of the entangled state. Our POVMs satisfy the usual relation that \begin{equation} \sum _i \hat{\Pi}_{\rm QA}^i=\hat{\bf 1}_4 \ . \end{equation} Alice then informs Rob of her measurement result and Rob performs one of four local operations, $\hat{B}^i_{\rm R}$, on his half of the joint system. After this protocol, Rob has the state $\rho _{\rm R}$  with a certain fidelity to Alice's teleported qubit, $F$. Provided that $| \psi \rangle _{\rm Q}$ is a pure state, $F$ can be quantified by the overlap ${}_{\rm Q} \langle \psi | \rho _{\rm R} | \psi \rangle _{\rm Q} ^2$. The fidelity of the protocol, $f$, is then calculated by averaging $\sqrt{F}$ over all possible values of $| \psi \rangle$. 

It is well known that the optimal value of $f$ that one can achieve with any given entangled state is given by the state's maximal overlap with a maximally entangled state \cite{horodecki1,horodecki2}. We begin our discussion by calculating this optimum for the entangled state $|\Psi \rangle _{\rm AR}$ given by Eq. (\ref{psipsi}). Since this is the state prepared by our initially inertial observers, the protocol that is optimal for this state is the protocol we will use when Rob undergoes uniform acceleration. That is, Rob's local operations are restricted to region I and is given by $\hat{B}^i_{I {\rm R}} \oplus \hat{\bf 1}_{I}$. 
Writing $| \Psi \rangle _{\rm AR}$ in the Schmidt basis,
\begin{equation} |\Psi \rangle _{AR} = \sum _{i\in \{0,1\}} \lambda _i | \varphi _i \rangle _{\rm A}| \vartheta _i \rangle _{\rm R} \ , \end{equation}
we can immediately derive the bound
\begin{equation} f \leq \left[ \frac{\lambda _0+\lambda _1}{\sqrt{2}} \right] \ . \end{equation}
It is easy to see that we saturate this bound using the following POVMs
\begin{eqnarray} \hat{\Pi}^1_{\rm QA} &=& \chi( |0\rangle |\varphi _0\rangle + |1 \rangle | \varphi _1 \rangle ) \ , \label{measure1} \\
\hat{\Pi}^2_{\rm QA} &=& \chi (|0\rangle |\varphi _0\rangle - |1 \rangle | \varphi _1 \rangle ) \ , \\
\hat{\Pi}^3_{\rm QA} &=& \chi ( |0\rangle |\varphi _1\rangle + |1 \rangle | \varphi _0 \rangle ) \ , \\
\hat{\Pi}^4_{\rm QA} &=& \chi ( |0\rangle |\varphi _1\rangle - |1 \rangle | \varphi _0 \rangle ) \ , \label{measure4} \end{eqnarray}
as well as the local operations
\begin{eqnarray} \hat{B}^1 _{\rm R} &=& |0 \rangle \langle \vartheta _0 | +|1 \rangle \langle \vartheta _1 | \ , \label{opy1} \\
\hat{B}^2 _{\rm R} &=& |0 \rangle \langle \vartheta _0 | -|1 \rangle \langle \vartheta _1 | \ , \\
\hat{B}^3 _{\rm R} &=& |1 \rangle \langle \vartheta _0 | +|0 \rangle \langle \vartheta _1 | \ , \\
\hat{B}^4_{\rm R} &=& |1 \rangle \langle \vartheta _0 | -|0 \rangle \langle \vartheta _1 | \ .\label{opy4}   \end{eqnarray}
We can then calculate Rob's final state, $\sigma _{\rm R}$, using the four POVMs (\ref{measure1})--(\ref{measure4}) and local operations $\hat{B}_{I \rm R}^i \oplus \hat{\bf 1}_{I}$ as defined by Eqs. (\ref{opy1})--(\ref{opy4}) on the state $|\psi \rangle _{\rm Q}\langle \psi|\otimes \rho _{I {\rm AR}}$ with $\rho _{I {\rm AR}}$ given by Eq. (\ref{effectiverho}).  The fidelity of the protocol is found numerically by taking the overlap of the $\sigma _{\rm R}$, with Alice's initial state and averaging this overlap over all possible values of $\alpha$ and $\beta$. To perform this calculation we take the initial value of $\alpha$ and $\beta$ to both be equal to $1/\sqrt{2}$. That is we take $|\psi \rangle _{\rm Q} =|+\rangle _{\rm Q}$  and apply a random unitary transformation, $\hat{U}$, which gives a final state $\sigma _{\rm R} (\hat{U}|+\rangle _{\rm Q})$. The fidelity is then
\begin{equation} f= \int d \nu (\hat{U}) \langle +| \hat{U}^\dagger \sigma _{\rm R} (\hat{U} |+\rangle) \hat{U} |+ \rangle \ . \end{equation}
Here we have omitted the subscript Q to avoid cumbersome notation. In Fig. 2 we calculate $f$ for $r=0.6$ and a sample of $2\times 10^5$ random unitary matrices. From this figure one can observe the remarkable result that the optimal choice of $\xi$ is non-zero implying that $f$ is maximized when one encodes in a non-orthogonal subspace.

\begin{figure}
\includegraphics[scale=0.43]{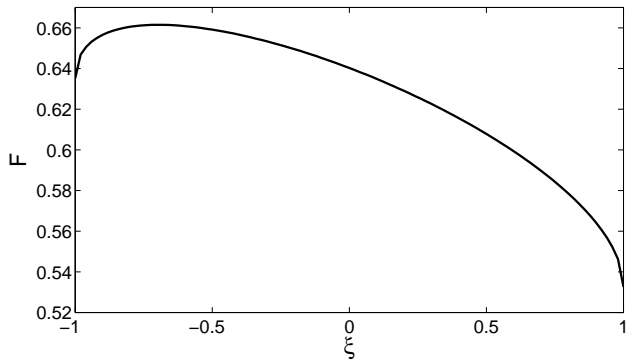}
\caption{\label{fig:epsart} The fidelity of Rob's final state as a function of $\xi $.  The value of the acceleration parameter is $r=0.6$.}
\end{figure}
\subsection{Distinguishability}
We now turn our attention to a simpler, but very instructive thought experiment. Once again Rob accelerates at a constant rate but this time is given two states. One state is the $|+\rangle _M$ state and the other is either $|+\rangle _ M$ or $|\varphi \rangle _M$. Rob then must determine whether his states are the same or different and perform  actions that depend on the result. Due to the acceleration, Rob's state is populated by the Unruh effect and his second state is either $\rho  _ I(|+\rangle)$ or $\rho  _I (|\varphi \rangle )$, which can be easily calculated using Eqs. (1),(2),(\ref{unruh1})--(\ref{unruh2}). If the states are different, by implementing the optimal measurement strategy, the probability that Rob can determine that the two states are indeed different is maximized when the following angle is maximal \cite{buresangle2} 
\begin{equation} \theta = \arccos \bigg( {\rm Tr} \big[ \sqrt{ \rho ^{(1)}_I(|+\rangle) ^{\frac{1}{2}} \rho ^{(2)} _I (|\varphi \rangle) \rho ^{(1)} _I (|+\rangle)^{\frac{1}{2}} }\big]  \bigg) \label{buresangle} \ .   \end{equation}
In Fig. 2 we calculate and plot $\theta$ as a function of $\xi$ for $r=0.85$. We see that $\theta$ is maximal for a non-zero value of $\xi$ implying once again that it is preferable to encode in non-orthogonal states.
\begin{figure}
\includegraphics[scale=0.43]{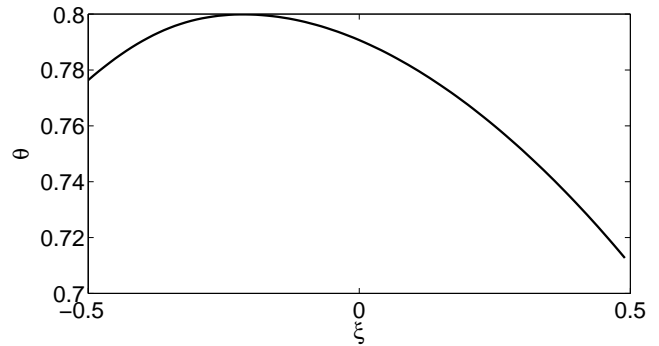}
\caption{\label{fig:epsart2} The probability that an accelerated observer can distinguish $|+ \rangle$ from $|\varphi \rangle$ is maximized by the maximal value of $\theta$. Therefore $\theta$ (in radians) is given as a function of $\xi$. Again $\xi$ is an orthogonality parameter for which $\xi =0$ denotes an orthogonal state in Minkowski space. The value of the acceleration parameter is $r=0.85$.  }
\end{figure}

\section{State space of the effective qubit}\label{three}
The last result in the previous section is highly suggestive as Eq. (\ref{buresangle}) is literally the Bures angle between two quantum states \cite{buresangle3}. The fact that the Bures angle between states changes in a way that cannot be attributed simply to a loss of purity alludes to the possibility that the state space occupied by the effective qubit may be deformed. In this section we show this analytically by calculating the state space metric for an effective qubit, $\rho _R$ as observed by an accelerating observer. There are a number of state space metrics proposed in the literature \cite{buresisgood}. However, the Bures metric is a metric that is both Riemannian and  monotonic and is therefore the most suitable for our analysis \cite{buresisgood}. The Bures metric is derived from the Bures distance which is given by \cite{burescalc,buresoriginal}
\begin{equation} D(\rho,\sigma) =2[1-F(\rho , \sigma)]  \label{buresmetric} \end{equation}
where
\begin{equation} F(\rho, \sigma ) = {\rm Tr}[ \sqrt{ \rho^\frac{1}{2} \sigma \rho ^\frac{1}{2}}]^2 \ . \label{fidelity} \end{equation}
It is then a simple matter of choosing $\sigma = \rho + d\rho $ to define the metric of the qubit's state space. We can then parameterize $\rho$ by the vector $\vec{n}=(x,y,z)$ to give a general expression for a qubit in Minkowski space as $(1,\vec{n}) {\bf \cdot } (1,\vec{\sigma}) /2$, where $\vec{\sigma}$ are the Pauli spin matrices. Similarly we can also write a general expression for $d\rho$ in Minkowski space as $\vec{dn} \cdot \vec{\sigma}$ where $\vec{dn}\equiv (dx,dy,dz)$. We can use these expressions for $\rho $ and $d\rho$ to derive the Bures metric of a general effective qubit as observed by an accelerating observer. However, since we know of no way to solve this problem for the full infinitely large density matrices that result from the Unruh effect, we content ourselves to derive the Bures metric for low accelerations. We henceforth ignore all terms of order $O(r^4)$ or smaller. This will turn out to be sufficient for our purposes. Using Eqs. (\ref{unruh1}) and (\ref{unruh2}), we write the density matrix of a general effective qubit subject to the Unruh effect in the limit of low accelerations,
\begin{equation} \rho \approx \frac{1}{2C^2} \left( \begin{array}{ccc} 1+z & \frac{x-iy}{C} & 0 \\ \frac{x+iy}{C} & \frac{(1-z)}{C^2}+T^2(1+z) & \frac{\sqrt{2}T^2 (x-iy)}{C} \\ 0 & \frac{\sqrt{2}T^2 (x+iy)}{C} & \frac{2T^2(1-z)}{C^2}   \end{array} \right) \ . \label{acceleratedqubit}  \end{equation}
Here and throughout the rest of this paper we use the shorthand that $C=\cosh r$ and $T=\tanh r$.  We can also define a general expression for $d\rho $ for small accelerations from Eq. (\ref{acceleratedqubit}) by simply replacing $1\pm z$ with $\pm dz$ and $\{x,y\}$ with $dx$ and $dy$ respectively. We note that our expression for $\rho$ does not have the property that  ${\rm Tr}[\rho] = 1$. This causes the complication that Eq. (\ref{buresmetric}) is not zero for $\sigma = \rho$. To solve this problem we make a slight generalization of the Bures distance for states that do not have a unity trace:
\begin{equation} D(\rho , \sigma )= 2[{\rm Tr}[\rho ] {\rm Tr}[\sigma ] - F(\rho , \sigma)] \label{generalbures} \ . \end{equation}
To be an appropriate extension of the Bures metric, our distance measure must satisfy the following five conditions: \cite{buresisgood,buresoriginal}
\begin{itemize}
\item The distance measure must be Riemannian.
\item The distance measure must be a monotone, that is $D(\rho, \sigma) \geq D\left({\cal E}( \rho ) , {\cal E}( \sigma)\right)$ for any completely positive trace preserving (CPTP) map, ${\cal E}$.
\item The distance measure must be symmetric in its outputs, that is $D(\rho , \sigma ) = D(\sigma , \rho )$.
\item The distance measure must be positive semi-definite, $D(\rho, \sigma) \geq 0$, with the bound only achieved for $\sigma = \rho$.
\item The distance measure must obey the triangle inequality, \begin{equation} D(\rho , \sigma) \leq D(\rho , \tau) +D(\sigma , \tau)  \ .\end{equation}

\end{itemize}
We will prove by construction that this modified distance measure is Riemannian (whereas the standard Bures distance is not for states with ${\rm Tr}[\rho ] <1$). That the second condition is satisfied immediately follows from the fact that the fidelity obeys monotonicity, $F(\rho, \sigma) \leq F\left({\cal E}( \rho ) , {\cal E}( \sigma)\right)$ and the fact that the function ${\rm Tr}[\rho] {\rm Tr} [\sigma]$ is non increasing under CPTP maps by definition. That the metric is still symmetric in its outputs is obvious from the fact that the new term is symmetric in its outputs. It is well known that the fidelity is maximized for $\sigma = \rho$ and that when this equality holds the fidelity can be written ${\rm Tr} [\rho]{\rm Tr} [\sigma]$ and therefore we see that the distance measure is indeed positive semi-definite. This brings us to the only non-trivial part of our proof that our distance measure is an appropriate modification of the Bures distance, that $D(\rho, \sigma)$ obeys the triangle inequality. To prove this we first note that the fidelity can be written $F(\sigma , \rho ) = \left(|| \sqrt{\rho} \sqrt{\sigma} ||\right)^2  \leq ||\rho|| \times || \sigma ||$ \footnote{See for instance the theorem in the appendix of K. M. R. Audenart, M. Nussbaum, A. Szkola, F. Verstrate, Commun. Math. Phys. {\bf 279}, 251 (2008) and set $s=1$.}. From this it immediately follows that
\begin{equation} F(\rho , \sigma ) \leq {\rm Tr}[\rho] {\rm Tr}[\sigma ] \label{tri1} \ . \end{equation}
The inequality we wish to prove is
\begin{eqnarray}   {\rm Tr} [\rho ] {\rm Tr} [\sigma] - F(\rho , \sigma ) &\leq & {\rm Tr} [\rho ] {\rm Tr} [\tau] - F(\rho , \tau ) \nonumber \\ &+& {\rm Tr} [\sigma ] {\rm Tr} [\tau] - F(\sigma , \tau ) \ . \nonumber \\ \end{eqnarray}
We first rearrange this inequality such that both sides are positive definite:
\begin{eqnarray} {\rm Tr} [\rho ] {\rm Tr} [\sigma] +F(\rho, \tau ) + F(\sigma , \tau ) &\leq& {\rm Tr} [\rho ] {\rm Tr} [\tau] +{\rm Tr} [\sigma ] {\rm Tr} [\tau] \nonumber \\ &+&  F(\rho , \sigma ) \ . \end{eqnarray}
Using Eq. (\ref{tri1}) we can then write
\begin{eqnarray} {\rm Tr} [\rho ] {\rm Tr} [\sigma] +F(\rho, \tau ) + F(\sigma , \tau ) &\leq & {\rm Tr} [\rho ] {\rm Tr} [\tau] +{\rm Tr} [\sigma ] {\rm Tr} [\tau] \nonumber \\ &+&  {\rm Tr} [\rho]  {\rm Tr} [\sigma ] \ ,  \end{eqnarray} which gives \begin{eqnarray}  F( \rho , \tau) + F(\sigma, \tau ) &\leq& {\rm Tr} [\rho ] {\rm Tr}[\tau ] +{\rm Tr} [\sigma ] {\rm Tr}[\tau ] \ , \nonumber \\ 
\end{eqnarray}
which is clearly satisfied by direct application of Eq. (\ref{tri1}).

Seeing that our modification to the Bures distance is appropriate we resume our derivation of the metric of the effective state space. From Eq. [\ref{acceleratedqubit}], it is easy to see that for $\sigma = \rho + d\rho$,  the product of the traces can be calculated 
\begin{equation} {\rm Tr}[\rho] {\rm Tr}[\rho + d\rho] \sim  1-2r^2(1-z)+r^2dz \ . \label{error} \end{equation}
To evaluate the Bures metric we begin by setting the radicand of Eq. (\ref{fidelity}) equal to $M$. That is, $M=\rho^{\frac{1}{2}}(\rho + d\rho)\rho^{\frac{1}{2}}$. The fidelity is given by 
\begin{equation} {\rm Tr}[M^\frac{1}{2}]^2 = (\sqrt{\lambda _1}+ \sqrt{\lambda _2} + \sqrt{\lambda _3})^2 \ , \label{radicand} \end{equation} where $\lambda _i$ are the eigenvalues of $M$. The characteristic equation for $M$ is
\begin{equation} \lambda^3 - {\rm Tr}[M] \lambda^2 + \alpha \lambda - {\rm Det}[M] = 0 \ .  \end{equation}
Here $\alpha$ is given by ${\rm Det}(M) {\rm Tr} [M^{-1}]$. Clearly the eigenvalues of $M$ remain the same if we redefine $M=\rho (\rho +d \rho)$. We can immediately simplify this equation by noting that the determinant of $M$ is of order $r^4$ and can therefore be ignored for our analysis. One can now easily show that, using this approximation, the characteristic equation reduces to
\begin{eqnarray} && \lambda ^2 - {\rm Tr}[\overline{\rho} (\overline{\rho}  + d\overline{ \rho})]\lambda + {\rm Det}[ \overline{\rho} (\overline{ \rho}   +d \overline{ \rho})] \nonumber \\ &\equiv & \lambda ^2 - {\rm Tr}[\overline{M}]\lambda +{\rm Det}[\overline{M}] \ , \end{eqnarray}
where
\begin{equation} \overline{\rho} = \frac{1}{2C^2} \left( \begin{array}{cc} 1+z & \frac{x-iy}{C} \\ \frac{x+iy}{C} & \frac{1-z}{C^2} + T^2(1+z) \end{array}  \right) \label{rhobar} \ .  \end{equation}
From Eq. (\ref{rhobar}) we can similarly find $\overline{\rho} + d\overline{\rho}$. Defining the eigenvalues of $\overline{M} $ as $\overline{\lambda}_1$ and $\overline{\lambda}_2$, we can write Eq. (\ref{radicand}) to second order in $r$:
\begin{eqnarray} {\rm Tr}[M^\frac{1}{2}]^2 &\sim & \overline{\lambda} _1 + \overline{\lambda} _2 +2 \sqrt{ \overline{\lambda}_1 \overline{\lambda}_2} \\ &=& {\rm Tr}[\overline{M}] + 2{\rm Det}[\overline{M}]^\frac{1}{2} \label{simple} \ .   \end{eqnarray}

We then can write the expression for the determinant of $\overline{M}$ as the product of the determinants of $\overline{\rho}$ and $\overline{\rho} + d\overline{\rho}$: 
\begin{eqnarray} {\rm Det}[\overline{\rho}] &=& \frac{1}{4C^2}\big[\frac{1-n^2}{C^2}+T^2(1+z)^2\big] \label{det1} \\ {\rm Det}[\overline{\rho} + d \overline{\rho}] &=& \frac{1}{4C^2}\left[\frac{1-n^2-2dS-dn^2}{C^2} \right. \nonumber \\ &+& \left. T^2(1+z+dz)^2 \right]    \label{det2} \end{eqnarray}
where $dS \equiv xdx+ydy+zdz$.
To evaluate ${\rm Det}[\overline{M}]^\frac{1}{2}$ it is useful to define
\begin{equation} G = \big[ 1 + T^2\frac{(1+z)^2}{1-n^2}\big] \ . \end{equation}
We can now express the determinant of $\overline{M}$ in terms of $G$ to second order in $r$ as follows:
\begin{eqnarray} {\rm Det}[\overline{M}] &\sim & \frac{1}{16C^8}\left[ (1-n^2)^2G^2-(2dS+dn^2) \right. \nonumber \\ &\times & \left. (1-n^2)G +  2(1-n^2)^2\frac{(G-1)}{(1+z)}dz \right. \nonumber \\ &+& \left. (1-n^2)^2\frac{(G-1)}{(1+z)^2}dz^2 \right] \ .  \label{factor} \end{eqnarray}
We then factor $(1-n^2)^2G^2$ out of Eq. (\ref{factor}), expand to second order and neglect terms of order $r^4$ or smaller to get the expression
\begin{eqnarray} {\rm Det}[\overline{M}]^\frac{1}{2} & \sim & \frac{1}{4C^4}\left[ (1-n^2)G^2- \frac{2dS+dn^2}{2}  \right. \nonumber \\ &+& \left. \frac{dS^2}{2(1-n^2)G} + T^2dz^2 +T^2dz(1+z) \right] \ . \nonumber \\ \label{nearly} \end{eqnarray}
By combining Eq. (\ref{simple}) with Eq. (\ref{nearly}), one can easily show that the fidelity, to second order in $r$, is given by 
\begin{eqnarray} F  &\sim & \frac{1}{4C^4}\left\{dn^2+T^2dz^2+\frac{dS^2}{(1-n^2)^2}[1-T^2\frac{(1+z)^2}{1-n^2}] \right\} \nonumber \\ &+& 1-2r^2(1-z)+r^2dz \ .  \label{nearlythere} \end{eqnarray} 
The last few terms in Eq. (\ref{nearlythere}) coincide exactly with Eq. (\ref{error}). We can therefore use Eq. (\ref{generalbures}) to cancel these terms and write the metric of the state space for an accelerated qubit in the limit of low acceleration as
\begin{equation} ds^2 = \frac{1}{4C^4}\left\{dn^2+T^2dz^2+\frac{dS^2}{1-n^2}[1-T^2\frac{(1+z)^2}{1-n^2}]\right\} \ . \label{buresfinal} \end{equation}
Note that Eq. (\ref{buresfinal}) differs in form for the Bures metric for a qubit in Minkowski space \cite{burescalc, buresisgood} proving that the Unruh effect deforms the state space of an effective qubit. Also it is easy to see that Eq. (\ref{buresfinal}) reduces to the metric of a sphere for $r=0$ as required. The metric takes on a particularly interesting form in polar coordinates, $(\xi, \theta, \phi)$ (where $\xi $ is a coordinate parameter not to be confused with our orthogonality parameter from before). We can write the metric tensor of our effective qubit, $\tilde{g}_{\mu \nu }$, as
\begin{equation} \tilde{g}_{\mu \nu } = \frac{1}{4C^4} \left(g_{\mu \nu} + h_{\mu \nu}\right) \ , \end{equation}
where $g_{\mu \nu }/4$ is the standard Bures metric of a qubit and $h_{\mu \nu }$ is a perturbation of order $r^2$. Explicitly they are
\begin{equation} g= \left[ \begin{array}{ccc} \frac{1}{1-\xi ^2} & 0 & 0 \\ 0 & \xi^2 & 0 \\ 0 & 0 & \xi ^2 \sin ^2 \theta  \end{array} \right]  \end{equation}
and
\begin{equation} h = \left[ \begin{array}{ccc} T^2 (1+\xi \cos \theta )^2 + \frac{T^2}{2}\cos ^2 \theta & - \frac{T^2 }{2}r \sin \theta \cos \theta & 0 \\ - \frac{T^2 }{2}r \sin \theta \cos \theta & \frac{T^2 \xi ^2}{2} \sin ^2 \theta & 0 \\ 0 & 0 & 0 \end{array} \right] \ . \end{equation}
Notice that while the original Bures metric of a qubit is homogenous and isotropic and thus has no preferred direction, this is not true of $h$.

Finally, to verify that the deformation of the effective state space is not just a coordinate change we calculate the scalar curvature to first order in $r^2$. Using the standard construction of Riemannian geometry and changing to polar coordinates, $(\xi,\theta,\phi)$ the scalar curvature is \begin{equation} R= (24 + \delta R )\cosh ^4 r \ , \end{equation} where \begin{eqnarray} \delta R &=& \frac{2T^2}{\xi ^2 (\xi ^2 -1)}\left[4+8 \xi ^2 - 15 \xi ^4 + 5 \xi ^6  \right. \nonumber \\ &-&\left. 8\xi(2\xi-3)\cos \theta (4+8\xi^2 - 11 \xi ^4 + 5 \xi ^6) \cos 2\theta  \right] \ . \nonumber \\   \end{eqnarray} The scalar curvature is no longer a constant but varies as a function of $\theta$ and $\xi$.

\section{conclusion}\label{four}
We have shown numerically, through the calculation of entanglement and two different quantum information protocols, properties that suggest that the structure of a qubit's state space changes when it is accelerated. This possibility was derived analytically for the small acceleration case. However, one actually needs to have a very large acceleration, perhaps produced near the event horizon of a black hole, to have an appreciable effect on the state space despite our low-$r$ approximation \cite{Milburnfollowup}. Note that we neglected the consideration of using quantum ancillary states for a phase reference. How reference states are affected by Unruh radiation and how the state space of a qubit acts in this scenario is an interesting topic for future research. Finally we note that this decoherence can be simulated by a two-mode squeezing operator \cite{Milburnfollowup, squeeze}. This implies that our results can be simulated in the laboratory and raises the question as to whether there are other decoherence maps that have similar effects on the state space. 
  % input acknowledgement

  The author wishes to thank Emre Guler and David Paganin for their thorough review of this manuscript as well as Daniel Terno, Josh Combes and Joan Vaccaro for some stimulating discussions. This research was funded in part by the Australian Research Council Center of Excellence for Particle Physics at the Tera-scale, an Australian Postgraduate Award, and a J. L. William Ph. D. scholarship.

\end{document}